\documentclass{PoS}
\usepackage{amsmath,bm}

\title{Three-nucleon contact interaction at the next-to-leading order}

\ShortTitle{3-nucleon contact interaction at NLO}

\author{\speaker{L. Girlanda}\\
        Dipartimento di Matematica e Fisica ``E. De Giorgi''\\
        Universit\`a del Salento, and\\
        INFN Sezione di Lecce, I-73100 Lecce, Italy\\
        E-mail: \email{girlanda@le.infn.it}}

\author{A. Kievsky\\
        INFN Sezione di Pisa, I-56127 Pisa, Italy\\
        E-mail: \email{kievsky@pi.infn.it}}

\author{M. Viviani\\
        INFN Sezione di Pisa, I-56127 Pisa, Italy\\
        E-mail: \email{viviani@pi.infn.it}}

\abstract{A satisfactory description of bound and scattering states of the three-nucleon (3N) system is still lacking, contrary to what happens in the two-nucleon case. In the framework of chiral effective theory, it is possible that a realistic 3N interaction will require the inclusion of subleading contact terms, which are unconstrained by chiral symmetry. We construct the minimal 3N contact Lagrangian imposing all constraints from the discrete symmetries, Fierz identities and Poincar\'e covariance, and show that it consists of 10 independent operators. By integrating out the pions from the effective theory we derive the pion-induced contributions to the corresponding low-energy constants (LECs), and find no clear signal of convergence of the pionful theory between N2LO and N3LO.}

\FullConference{Sixth International Conference on Quarks and Nuclear Physics,\\
		April 16-20, 2012\\
		Ecole Polytechnique, Palaiseau, Paris}

\begin{document}

\section{Introduction}
One of the central tasks in nuclear physics is the determination of the interaction among nucleons. Traditionally, the two-nucleon ($2N$) forces have been modeled phenomenologically. By now, these models reach a remarkable degree of accuracy, providing an excellent fit to the very large body of experimental data. More recently, so called chiral potentials have been developed, which establish a solid link between the nuclear interaction and the underlying strong interaction theory \cite{bernard95,vankolck99,bedaque02,epelbaum06,epelbaum09}. They are derived in the framework of an effective field theory, incorporating all symmetry properties of QCD, and organized as a systematic expansion in powers of momenta and quark masses.
This progress, together with the tremendous improvement in ab-initio numerical calculations  of bound and scattering states of few-nucleon systems \cite{leidemann12}, make the latter a privileged theoretical laboratory to test our understanding  of the nuclear interaction.

Presently  the $2N$ potential has been worked out up to the next-to-next-to-next-to leading order (N3LO) in the low-energy expansion \cite{entem03,epelbaum04}, yielding a $\chi^2$ per datum very close to one. In the three-nucleon ($3N$) sector, however, the situation is not as satisfactory. A $3N$ interaction starts to show up at the N2LO \cite{vankolck94,epelbaum02} in the effective theory. From the phenomenological point of view, its contribution is necessary in order to reproduce the 3- and 4-nucleon binding energies.
Nevertheless, all currently available $3N$ potentials fail to simultaneously describe also the neutron-deuteron doublet scattering length \cite{kievsky10}. As a matter of fact, several low-energy data in the nucleon-deuteron system remain still unexplained, the most prominent one being the so-called $A_y$ puzzle. 

It is interesting to notice that the discrepancies between theory and experiment in the $3N$ system arise especially at very low energy, and disappear as the energy increases. At such low energies, the nuclear interactions are effectively point-like. We therefore propose to refine that part of the interaction which stems from purely nucleonic vertices.  Indeed, at very low energies, even the pions can be integrated out of the theory, giving rise to the ``pionless" effective theory. The inclusion of contact interactions, which are unconstrained by chiral symmetry, could also provide the necessary flexibility to describe the $3N$ interaction. In this respect it is fitting to recall that all adopted $3N$ interaction models only contain a few free parameters, contrary to what happens for the $2N$ interaction, which is parametrized by more than 20 adjustable parameters. For example, the chiral $3N$ force at N2LO only contains 2 adjustable LECs, and its extension to N3LO does not involve any new LEC \cite{bernard08,bernard11}.

 Invariance under parity dictates that subleading $3N$ contact terms contain two (covariant) derivatives of nucleon fields. In the pionful version of the effective theory such terms would therefore contribute at N4LO, and are therefore beyond the accuracy of present day models, whereas in the pionless theory they represent the first corrections to the $3N$ interaction. It may well be that, at very low energy, the latter expansion scheme is more effective.
 Our task is twofold: first, classify all possible terms of this kind, second, fit the accompanying LECs to $3N$ data, in association with a given $2N$ potential model. We report here our progress in the first task. 
Contact terms  have to respect the discrete symmetries of the underlying theory, namely (besides parity) charge conjugation, or better to say, in a non-relativistic theory, time-reversal invariance. In addition, strong constraints come from Fierz identities combined with the anticommutation properties of field operators, or equivalently, from the fact that the nuclear wave function is antisymmetric under exchange of two particles. As an illustration we consider in the next section the leading (no derivatives) isospin-breaking contact operators.

\section{Leading isospin-breaking contact operators}
The most general isospin-violating, charge-invariant isospin structures of nucleons $i,j,k$ are 
\begin{equation}
\begin{array}{l}                
T^+= \tau_i^3, \quad  ({\bm \tau}_i \cdot {\bm \tau}_j) \tau_k^3, \quad i\neq\j\neq k,\\
T^-= ({\bm \tau}_i \times {\bm \tau}_j )^3,
\end{array}
\end{equation}
where the superscript refer to parity under time reversal.
Even (odd) isospin structures have to be associated with (rotation invariant) spin structures containing even (odd) numbers of $\sigma$ matrices.
As a result, a list of seven operators is found
\begin{eqnarray}
O_1^\tau &=& N^\dagger \tau^3 N N^\dagger N N^\dagger N,\\
O_2^\tau &=& N^\dagger \tau^3 \sigma^i N N^\dagger \sigma^i N N^\dagger N,\\
O_3^\tau &=& N^\dagger \tau^3  N N^\dagger \sigma^i N N^\dagger \sigma^iN,\\
O_4^\tau &=& \epsilon^{ab3} \epsilon^{ijk} N^\dagger \tau^a  \sigma^i N N^\dagger \tau^b \sigma^j N N^\dagger \sigma^k N,\\
O_5^\tau &=& N^\dagger \tau^a  N N^\dagger \tau^a  N N^\dagger \tau^3 N,\\
O_6^\tau &=& N^\dagger \tau^a  \sigma^i N N^\dagger \tau^a  \sigma^i N N^\dagger \tau^3 N,\\
O_7^\tau &=& N^\dagger \tau^a  \sigma^i N N^\dagger \tau^a   N N^\dagger \tau^3 \sigma^iN.
\end{eqnarray}
Simultaneous exchanges of spin and isospin indices allow to write a set of linear relations among these operators. For example, from the exchange of nucleon fields 1-2,
\begin{eqnarray}
O_1^\tau &=& -\frac{1}{2} O_1^\tau - \frac{1}{2} O_2^\tau,\\
O_2^\tau &=& -\frac{3}{2} O_1^\tau + \frac{1}{2} O_2^\tau,\\
O_3^\tau &=& -\frac{1}{2} O_2^\tau -\frac{1}{2} O_3^\tau - \frac{1}{4} O_4^\tau,\\
O_4^\tau &=& 2 O_2^\tau - 2 O_3^\tau,\\
O_5^\tau &=& -\frac{3}{4} O_1^\tau -\frac{3}{4} O_3^\tau + \frac{1}{4} O_5^\tau + \frac{1}{4} O_6^\tau,\\
O_6^\tau &=& -\frac{9}{4} O_1^\tau + \frac{3}{4} O_3^\tau + \frac{3}{4} O_5^\tau - \frac{1}{4} O_6^\tau,\\
O_7^\tau &=& -\frac{3}{2} O_2^\tau + \frac{1}{2} O_7^\tau.
\end{eqnarray}
Two further set of relations are found by exchanging the other two pairs of nucleons. It turns out that six relations are linearly independent, which implies that only one operator survives, for example $O_1^\tau$.

\section{Subleading isospin-conserving contact operators}

The same procedure can be applied to all possible isospin-invariant $3N$ contact operators with two spatial derivatives \cite{girlanda11}. The initial list of 146 operators is reduced to 14 after imposing the Fierz constraints, which in this case also involve the field derivatives. A further reduction can be operated invoking Poincar\'e covariance, which can be implemented order by order in the low-energy expansion, as done in \cite{girlanda10}. At the order we are working, this requirement amounts to select operators which do not involve the total momentum \footnote{We ignore for the time being the ``fixed terms'', whose couplings are determined in terms of the lowest-order LECs.}.  As a result, a list of 10 independent operators is found, and the $3N$ contact potential in momentum space can be written as
\begin{eqnarray}
V=\sum_{i\neq j\neq k} &&\biggl[ -E_1 {\bf k}_i^2 - E_2 {\bf k}_i^2 {\bm
  \tau}_i\cdot {\bm \tau}_j - E_3{\bf k}_i^2 {\bm
  \sigma}_i\cdot {\bm \sigma}_j - E_4 {\bf k}_i^2 {\bm
  \sigma}_i\cdot {\bm \sigma}_j {\bm
  \tau}_i\cdot {\bm \tau}_j
\nonumber \\
&& - E_5 \left( 3 {\bf k}_i \cdot {\bm \sigma}_i {\bf k}_i\cdot {\bm
  \sigma}_j - {\bf k}_i^2\right) -E_6 \left( 3 {\bf k}_i \cdot {\bm \sigma}_i {\bf k}_i\cdot {\bm
  \sigma}_j - {\bf k}_i^2\right)) {\bm
  \tau}_i\cdot {\bm \tau}_j
\nonumber \\
&&+\frac{i}{2} E_7  {\bf k}_i \times \left({\bf Q}_i - {\bf Q}_j\right) \cdot ({\bm \sigma}_i +
          {\bm \sigma}_j) +\frac{i}{2} E_8  {\bf k}_i \times \left({\bf Q}_i - {\bf Q}_j\right) \cdot ({\bm \sigma}_i +
          {\bm \sigma}_j) {\bm  \tau}_j\cdot {\bm \tau}_k
\nonumber \\
&& - E_9 {\bf k}_i \cdot {\bm \sigma}_i {\bf k}_j \cdot {\bm \sigma}_j
- E_{10} {\bf k}_i \cdot {\bm \sigma}_i {\bf k}_j \cdot {\bm \sigma}_j
{\bm  \tau}_i\cdot {\bm \tau}_j \biggr],
\end{eqnarray}
where ${\bf k}_i = {\bf p}_1 - {\bf p}_i'$ and ${\bf Q}_i= {\bf p}_1 + {\bf p}_i'$ are given in terms of the initial and final momenta of nucleon $i$, respectively ${\bf p}_i$ and ${\bf p}_i'$.
Using a momentum cutoff depending only on momentum transfers, as done in \cite{navratil07}, e.g. $F({\bf k}_j^2; \Lambda) F({\bf k}_k^2; \Lambda)$, the ensuing coordinate space $3N$ potential has a local form, in terms of the function 
\begin{equation}
Z_0(r;\Lambda) = \int \frac{d{\bf p}}{(2 \pi)^3} {\mathrm{e}}^{i {\bf p}\cdot {\bf r}} F({\bf p}^2; \Lambda)
\end{equation}
and its derivatives. Explicitly,
\begin{eqnarray}
V=\sum_{i\neq j\neq k} && (E_1 + E_2 {\bm \tau}_i \cdot {\bm \tau}_j + E_3 {\bm \sigma}_i \cdot {\bm \sigma}_j + E_4 {\bm \tau}_i \cdot {\bm \tau}_j  {\bm \sigma}_i \cdot {\bm \sigma}_j)  \left[ Z_0^{\prime\prime}(r_{ij}) + 2 \frac{Z_0^\prime(r_{ij})}{r_{ij}}\right] Z_0(r_{ik})  \nonumber \\
&& + (E_5 +E_6 {\bm \tau}_i\cdot{\bm \tau}_j) S_{ij} \left[ Z_0^{\prime\prime}(r_{ij}) - \frac{Z_0^\prime(r_{ij})}{r_{ij}}\right] Z_0(r_{ik}) \nonumber \\
&& + (E_7 + E_8 {\bm \tau}_i\cdot{\bm \tau}_k) ( {\bf L}\cdot {\bm S})_{ij} \frac{Z_0^\prime(r_{ij})}{r_{ij}} Z_0(r_{ik}) \nonumber \\
&& + (E_9 + E_{10} {\bm \tau}_j \cdot {\bm \tau}_k) {\bm \sigma}_j \cdot \hat {\bf r}_{ij}  {\bm \sigma}_k \cdot \hat {\bf r}_{ik} Z_0^\prime(r_{ij}) Z_0^\prime(r_{ik})
\end{eqnarray}
where $S_{ij}$ and $ ( {\bf L}\cdot {\bm S})_{ij}$ are respectively the tensor and spin-orbit operators for particles $i$ and $j$. 
As it is apparent, a choice of basis has been made such that most terms in the potential can be viewed as an ordinary interaction of particles $ij$ with a further dependence on the coordinate of the third particle. In particular, the terms proportional to $E_7$ and $E_8$ are of spin-orbit type, and, as already suggested in the literature \cite{kievsky99}, have the right properties to solve the $A_y$ puzzle (see also Ref.~\cite{canton00,canton01}).

\section{Matching to the pionful theory}

In the pionful theory the $3N$ interaction starts at N2LO, then it is parameter-free at N3LO \cite{bernard08,bernard11}, and it's being currently investigated at N4LO \cite{krebs12}. In the very low-energy regime $p\ll M_\pi$ pion exchange diagrams reduce to $3N$ contact terms. 
This is achieved by Taylor-expanding the corresponding $3N$ potential in momentum space. For example, the two-pion exchange component of the $3N$ potential at N2LO, reduce at low energy, to
\begin{equation}
V_{2\pi}^{\mathrm{N2LO}} \sim - \frac{g_A^2 c_1}{ 2 F_\pi^4 M_\pi^2 } {\bm \sigma}_1\cdot {\bf k}_1  {\bm \sigma}_2\cdot {\bf k}_2 {\bm \tau}_1 \cdot {\bm \tau}_2 + {\mathrm{perm.}} +O(k^4), 
\end{equation}
while the one-pion exchange component yields
\begin{equation}
V_{1\pi-{\mathrm{cont}}}^{\mathrm{N2LO}} \sim -\frac{g_A D}{8 F_\pi^2 M_\pi^2 } {\bm \sigma}_1\cdot {\bf k}_2  {\bm \sigma}_2\cdot {\bf k}_2 {\bm \tau}_1 \cdot {\bm \tau}_2 + {\mathrm{perm.}}  + O(k^4).
\end{equation}
Afterwards, the induced operators have to be expressed in the chosen basis, using the explicit relations  provided in \cite{girlanda11}.
An analogous procedure can be applied to the N3LO expressions provided in Refs.~\cite{bernard08,bernard11}, in a straightforward, although tedious manner. As a result, one gets the pion-induced contributions to the $E_{1,...,10}$ as functions of the lowest order LECs,
\begin{eqnarray}
E_1 &=& \frac{755 g_A^6}{24576 \pi F_\pi^6 M_\pi} + \frac{g_A^4}{256 \pi F_\pi^6 M_\pi} - \frac{g_A^4 C_T}{64 \pi F_\pi^4 M_\pi} - \frac{g_A^2 C_T}{8 m F_\pi^2 M_\pi^2} \sim {{0.10}}, \\
E_2  &=&  \frac{601 g_A^6}{36864 \pi F_\pi^6 M_\pi} + \frac{23 g_A^4 C_T}{384 \pi F_\pi^4 M_\pi} - \frac{5 g_A^2 C_T}{192 \pi F_\pi^4 M_\pi} - \frac{g_A^2 (5 C_T + 2 C_S)}{48 m F_\pi^2 M_\pi^2}  \sim {{0.06}}, \\
E_3 &=&  -\frac{3 g_A^6}{2048 \pi F_\pi^6 M_\pi} + \frac{3 g_A^4 C_T}{64 \pi F_\pi^4 M_\pi} + \frac{9 g_A^2  C_T }{16 m F_\pi^2 M_\pi^2} \sim {{0.00}},\\
E_4 &=&  -\frac{ g_A^6}{1024 \pi F_\pi^6 M_\pi} - \frac{3 g_A^2 C_T}{16 m F_\pi^2 M_\pi^2}  \sim {{0.00}}, \\
E_5 &=& \frac{79 g_A^6}{12288 \pi F_\pi^6 M_\pi} + \frac{g_A^4}{256 \pi F_\pi^6 M_\pi} - \frac{g_A^4 C_T}{64 \pi F_\pi^4 M_\pi} - \frac{g_A^2 C_T}{8 m F_\pi^2 M_\pi^2}  \sim {{0.02}},\\
E_6 &=& \frac{319 g_A^6}{36864 \pi F_\pi^6 M_\pi} + \frac{g_A^4}{256 \pi F_\pi^6 M_\pi} - \frac{g_A^2 (C_S -2 C_T)}{24 m \pi F_\pi^2 M_\pi^2}  \sim {{0.04}},\\
E_7 &=& -\frac{83 g_A^6}{6144 \pi F_\pi^6 M_\pi} - \frac{3 g_A^4}{128 \pi F_\pi^6 M_\pi} + \frac{3 g_A^2 C_T}{4 m F_\pi^2 M_\pi^2} \sim {{-0.08}},\\
E_8 &=&  -\frac{7 g_A^6}{3072 \pi F_\pi^6 M_\pi} - \frac{ g_A^4}{128 \pi F_\pi^6 M_\pi} + \frac{ g_A^2 C_T}{4 m F_\pi^2 M_\pi^2} \sim {{-0.02}},\\
E_9 &=& \frac{193 g_A^6}{4096 \pi F_\pi^6 M_\pi}  - \frac{ 3 g_A^2 C_T}{8 m F_\pi^2 M_\pi^2}  \sim {{0.14}} ,\\
E_{10} &=&  {{\frac{c_1 g_A^2}{2 F_\pi^4 M_\pi^2} + \frac{g_A D}{8 F_\pi^2 M_\pi^2}}}  +\frac{427 g_A^6}{12288 \pi F_\pi^6 M_\pi} + \frac{9 g_A^4}{512 \pi F_\pi^6 M_\pi} - \frac{ g_A^2(C_S+ C_T)}{8 m F_\pi^2 M_\pi^2} \sim {{0.09}}. 
\end{eqnarray}
The (rough) numerical values are given in units of $F_\pi^4 M_\pi^3$. Notice that only $E_{10}$ receives contributions from both N2LO ($\sim -0.05$) and N3LO ($\sim 0.15$) and there is no sign of convergence.  At N4LO there will appear the ``genuine'' contact contributions: if the convergence pattern is so bad, it is possible that they be phenomenologically relevant. In any case, the choice of a purely contact $3N$ interaction (including the subleading one), with a cutoff $\sim M_\pi$, could turn out to be rather effective from the practical point of view.

\section{Outlook}
In view of the above remarks, although in the chiral theory the subleading contact terms would only contribute at N4LO, they could play an important role, at sufficiently low energy, and provide the necessary flexibility to improve the description of the scattering observables, in particular the $N-d$ vector analyzing powers $A_y$ and $i T_{11}$ and the tensor analyzing power $T_{21}$, as well as several breakup cross sections \cite{sagara10}. Having fitted, in conjonction with a given realistic two-nucleon interaction, these $3N$ observables, the single $4N$ contact term \cite{girlanda11} could be used to reproduce the $^4$He binding energy, and the so-determined interactions tested in four-body scattering states.

\end{document}